# Superconductivity in Bismuth Oxysulfide $Bi_4O_4S_3$

Clastin I. Sathish,[1,2,*] Hai Luke Feng,[1,2] Youguo Shi,[3] and Kazunari Yamaura [1,2,#]

[1] Superconducting Properties Unit, National Institute for Materials Science, 1-1 Namiki, Tsukuba, Ibaraki 305-0044, Japan

[2] Graduate School of Chemical Sciences and Engineering, Hokkaido University, Sapporo, Hokkaido 060-0810, Japan

[3] Institute of Physics, Chinese Academy of Sciences, Beijing 100190, China

Bismuth oxysulfide $Bi_4O_4S_3$, which has recently been claimed to be an exotic superconductor ($T_c$ = 4.5 K), was investigated by magnetic susceptibility and electrical resistivity measurements as well as by electron probe microanalysis. Single-phase $Bi_4O_4S_3$ was successfully prepared by a high-pressure method, and its lattice parameters and normal-state resistivity, as well as the density of states at the Fermi level, were found to be comparable to those determined earlier. However, the observed superconductivity was most likely impurity-driven, strictly contradictory to the observations in ongoing experiments. The present results indicate that the superconductivity of $Bi_4O_4S_3$ does not truly reflect the bulk nature of the $BiS_2$ layered phase, regardless of the manner in which $Bi_4O_4S_3$ is synthesized. We discuss possible superconducting impurities.

[*] E-mail: CLASTIN.Sathish@nims.go.jp
[#] E-mail: YAMAURA.Kazunari@nims.go.jp

## 1. Introduction

Very recently, several independent research groups in Japan, China, and India have reported that bismuth oxysulfide $Bi_4O_4S_3$ shows exotic bulk superconductivity (SC) in the temperature range of 4.4–8.6 K.[1–3] The $BiS_2$ layer (square-planar) has been claimed to be the main factor inducing SC in $Bi_4O_4S_3$.[4,5] Although the observed superconducting transition temperatures ($T_c$) are relatively lower than those of other high-$T_c$ materials, this compound is a potential candidate for many applications in the field of physics because high-$T_c$ SC is observed only in a layered structure. In addition, Hall coefficient measurements suggest multiband SC in $Bi_4O_4S_3$, as in the case of Fe-based superconductors.[5] Hence, further investigation of the SC of $Bi_4O_4S_3$ may facilitate greater understanding of the essential physics related to high-$T_c$ SC.

Characterization of a polycrystalline sample of $Bi_4O_4S_3$ indicated that the lower and upper critical fields ($H_{c1}$ and $H_{c2}$) are approximately 15 Oe and 31 kOe,[3] respectively, but a much smaller $H_{c1}$ of ~7.0 Oe has been reported in another study.[1] However, regardless of the small variations in the superconducting properties of $Bi_4O_4S_3$, ongoing studies reveal a large superconducting shielding fraction and zero resistivity. $Bi_4O_4S_3$ crystallizes in a tetragonal structure with the $I4/mmm$ space group (at room temperature). There is a slight discrepancy in the lattice parameters reported for this compound: $a$ = 3.9592(1) Å and $c$ = 41.241(1) Å in ref. 1; $a$ = 3.978 Å and $c$ = 41.07 Å in ref. 2; and $a$ = 3.9697(2) Å and $c$ = 41.3520(1) Å in ref. 3. Because the above-mentioned differences between the values of $a$ and $c$ are less than 0.5% and 0.7%, respectively, it is difficult to establish a reasonable correlation among the lattice parameters and the reported superconducting properties.

To reveal the comprehensive SC properties of $Bi_4O_4S_3$, we attempted to grow a high-quality single crystal of $Bi_4O_4S_3$. We adopted a high-pressure (HP) method because the extremely low melting points of Bi (272 °C; boiling point: 1564 °C) and S (115.2 °C; boiling point: 444.7 °C) render crystal growth by the conventional methods at ambient pressure (AP) difficult. Although a high-quality single crystal has not yet been grown, we surprisingly found that the polycrystalline $Bi_4O_4S_3$ known to date shows most likely impurity-driven SC, in sharp contrast to the observations made in ongoing studies.  Although the compound turns out to be SC with zero resistivity on cooling to 1.5 K, SC has only a 0.3% shielding fraction at most in magnetization (100% is expected for an ideal bulk SC), indicating that the SC unlikely reflects the bulk nature of the $BiS_2$ layered phase.  An electron probe microanalyzer (EPMA) suggested that a Bi-rich and S-poor compound, which is possibly amorphous, might be responsible for the observed SC.

We also found that a similar superconducting impurity might complicate the ongoing studies on SC of the $Bi_4O_4S_3$ compound prepared by a normal process because its full shielding fraction was markedly decreased to less than 1/14 of that of the pellet even upon gentle grinding.  The London penetration depth ($\lambda_L$) and relative sample size unlikely accounted for the marked reduction.  This was supported by the fact that even a much weaker applied magnetic field of 1 Oe ($<<H_{c1}$) did not improve the reduced shielding fraction.  Instead, the mechanical break of a superconducting link thinly spread throughout the grain boundaries and at the surface in a synthesized pellet more reasonably accounted for the marked decrease.  In the present paper, we report the results of more detailed measurements of $Bi_4O_4S_3$ compounds, which were synthesized in the HP and normal processes, under various conditions using both bulk pellet and powder forms.

## 2. Experimental

Polycrystalline $Bi_4O_4S_3$ was prepared from fine $Bi_2O_3$ (99.99%; Soekawa Chemical Co., Tokyo), $Bi_2S_3$ (99.9%; Kojundo Chemical Lab. Co., Sakado), and S (99.99%; Kojundo Chemical Lab. Co.) powders.  The above-mentioned powders were mixed in a stoichiometric molecular ratio and placed in a cubic multianvil HP apparatus.  Then, they were heated at a specific temperature between 500 and 900 °C for 30 min at an elevated pressure of 3 GPa and quenched to room temperature before releasing the pressure.  The temperature was monitored by a W:Re5-W:Re26 thermocouple during heating.  The mixture was sealed in a Pt capsule and separated from Pt by a hexagonal boron nitride inner cell.  The inner cell did not react with the mixture components, as was evident from its unchanged color (white) after heating.  For reference, Bi powder (99.9%; Kojundo Chemical Lab. Co.) was heated under the same conditions at 700 °C for 30 min.  For comparison, polycrystalline $Bi_4O_4S_3$ was prepared by the normal method reported previously.[1–3] The stoichiometric mixture was pressed into a pellet and heated at 450 °C for 10 h in an evacuated quartz tube.

All the as-prepared compounds were investigated by X-ray diffraction (XRD) analysis under ambient conditions in an X'Pert Pro system (PANalytical) using monochromatic Cu-K$\alpha$ radiation. Selected compounds were subjected to XRD Rietveld analysis using RIETAN-FP software [6] and EPMA in JXA-8500F from JEOL. The XRD data for structure refinement were recorded in a RIGAKU X-ray diffractometer at room temperature ranging from 10–120° in $2\theta$ at a step interval of 0.03° using Cu-Ka radiation. The current and voltage needed to generate the X-ray were 40 mA and 40 kV, respectively. Magnetic susceptibility ($\chi$) measurements were performed using a magnetic property measurement system (MPMS, Quantum Design) in the temperature range of 2–300 K and at an applied magnetic field of 1 Oe or 10 Oe or 10 kOe, under field-cooling (FC) and zero-field-cooling (ZFC) conditions. The electrical resistivity ($\rho$) of the polycrystalline pellets was measured by a physical property measurement system (PPMS, Quantum Design) at temperatures between 2 and 300 K. The conventional four-terminal method was used for this purpose; the ac-gauge current and frequency were 1 or 2 mA and 110 Hz, respectively. Silver epoxy was used to fix gold wires onto the pellets for the $\rho$ measurements. Additional measurements for $\chi$ and $\rho$ were conducted down to 1.6 and 1.4 K in an MPMS VSM (Quantum Design) and a $^3$He probe in PPMS, respectively, for a piece of a selected sample pellet.

## 3. Results and Discussion

All the major XRD peaks for the sample prepared at 700 °C were assigned by assuming a tetragonal unit cell with the *I*4/*mmm* space group (Fig. 1), as in other ongoing studies.[1–3] The lattice parameters $a$ = 3.9592(1) Å and $c$ = 41.241(1) Å were comparable with the previously reported values. When the heating temperature was increased to 900 °C, a different XRD profile was observed, indicating the transformation of the tetragonal phase into an unidentified phase. On the other hand, the compounds synthesized at temperatures lower than 700 °C contained a small amount of the tetragonal phase and Bi metal and multiple impurities. In the present study, 700 °C was identified as the nearly optimal temperature for synthesizing polycrystalline $Bi_4O_4S_3$ at 3 GPa, while 500–510 °C was the optimal temperature range at AP.[1–3] Interestingly, the same XRD profile was obtained even after a week for the powder synthesized at 3 GPa, suggesting that $Bi_4O_4S_3$ synthesized by the HP method is chemically stable under ambient conditions, as is the powder synthesized by the normal process.[1–3]

To refine the structure parameters of the HP compound $Bi_4O_4S_3$, Rietveld analysis was conducted, and the results are summarized in Table I and Fig. 2(a). The $R$ factors for the refinements imply that the structure model mentioned above is indeed reasonable for the HP compound $Bi_4O_4S_3$. Meanwhile, the results indicate that small amounts of impurities are mixed in the compound. Unfortunately, we were unable to identify the impurities in this study. In the refinement, the S3 and O2 sites were assumed to be partially occupied and fixed to the $Bi_4O_4S_3$ composition.[3] Note that the O2 coordinate parameters $y$ and $z$ could not be refined because only a quarter of the O2 site was occupied; hence the parameters $y$ and $z$ of O2 were fixed to those obtained in a preliminary refinement with the assumption that the O2 occupancy factor is 0.5. We also tested a similar structure model proposed in Ref. 1, where the site occupancies of S3 and O2 were fixed to be 1 and 0.5, respectively. The alternative model, however, did not help to improve the refinement quality. Perhaps, a neutron diffraction study can refine the structure parameters further.

Figures 2(b) and 2(c) show a comparison of the coordination environment of Bi in the $BiS_2$ layer between the $Bi_4O_4S_3$ phases prepared by the HP method and a normal method. A significant difference is not obvious; however, the length of the S1–Bi2 bond, which is perpendicular to the $BiS_2$ layer, is approximately 4% shorter in the HP phase than that in the AP phase (2.66 and 2.77 Å, respectively). Note that the interlayer distances between the $BiS_2$ layers were 3.19(1) Å for the HP-prepared $Bi_4O_4S_3$ and 3.24(1) for the normally prepared $Bi_4O_4S_3$ (ref. 1), indicating a trivial change.

The superconducting properties of $Bi_4O_4S_3$ prepared by the normal method were studied for reference. Magnetic measurements clearly showed a superconducting transition with a notable shielding fraction (~129%, see the ZFC curve in Fig. 3) (the density of the compound was 6.11 g/cm$^3$, which was measured), as reported in ongoing studies.[1-3] The slight overestimation of the fraction was probably due to the demagnetizing effect, because the measurement was conducted on a sintered pellet. The $\rho$ vs $T$ curve for the polycrystalline $Bi_4O_4S_3$ pellet (inset of Fig. 3) shows a metallic character in the normal state and a sharp drop to zero, as reported previously.[1-3] Thereafter, the pellet was gently ground to an amount of loose powder using an agate mortar and pestle and re-subjected to magnetic measurements. Noticeably, the shielding fraction decreased to less than 1/14 of that of the pellet. To our best knowledge, the radical decrease in the shielding fraction has not been mentioned in the literatures thus far.[1-3] Note that the chemical and mechanical instabilities of $Bi_4O_4S_3$ were unlikely responsible for the radical decrease because the XRD pattern was essentially unchanged over the measurements. The radical decrease seemingly indicated that a superconducting link was physically destroyed by grinding. We discuss a possible origin of the radical decrease later.

Figure 4 shows the $\chi$–$T$ data measured at a weak magnetic field of 10 Oe for the compounds synthesized under HP conditions in a variety of heating temperatures. The powders used in the XRD measurements were reused in the magnetic susceptibility measurements. The desired amount of each loose powder sample was set in a sample holder and subjected to the $\chi$–$T$ measurements. The sample prepared at 650 °C showed a notable diamagnetic transition, presumably a superconducting transition, at 4 K. However, the superconducting shielding fraction was less than 2.2% at the low-temperature limit (the density of the compound was assumed to be 7.84 g/cm$^3$, which corresponded to the calculated density). Note that identical measurements of a pellet (instead of the powder) revealed a comparable shielding fraction, being in stark contrast to the full shielding fraction reported for compounds prepared by the normal process in other ongoing studies.[1–3]

Because the compound prepared at 650 °C contained a small amount of $Bi_4O_4S_3$, we investigated the rest of the compounds in the same manner to confirm the reported SC for $Bi_4O_4S_3$. Surprisingly, the nearly single phase $Bi_4O_4S_3$ prepared at 700 °C showed negligible anomaly in the $\chi$–$T$ curves, similarly to the reference material of Bi and the compound prepared at 900 °C. Thus, the magnetic data suggested the absence of the bulk SC in $Bi_4O_4S_3$ synthesized at 3 GPa. Unfortunately, a systematic correlation between the superconducting properties and the synthesis conditions was hardly established, and the superconducting volume fraction was just 2.2% at best in the HP-synthesized samples. It was therefore highly challenging to identify the phase responsible for the observed SC in the HP-synthesized samples.

The inset of Fig. 4 shows the $\chi$–$T$ curves measured under a much stronger magnetic field of 10 kOe for the near-single-phase $Bi_4O_4S_3$ (synthesized at 700 °C). The curves were almost temperature-independent, and a small Curie–Weiss-like upturn, which was indicative of an impurity contribution, was found in the low-temperature limit. The upturn part was much smaller than that reported in previous studies of $Bi_4O_4S_3$,[2] indicating the improved quality of the present compound. Furthermore, the temperature-independent $\chi_0$ was comparable to that determined in an ongoing study ($0.736 \times 10^{-3}$ emu mol$^{-1}$ Oe$^{-1}$).[2] Because $Bi_4O_4S_3$ is metallic in nature, the aforementioned similarity in $\chi_0$ implied that the densities of states at the Fermi level were comparable in the two cases.

For further characterization, the $\rho$–$T$ curve of the nearly single phase $Bi_4O_4S_3$ (synthesized at 700 °C) was measured (Fig. 5) down to 1.4 K. The data revealed that the compound shows metallic behavior over the measured temperature range; the room-temperature $\rho$ was approximately 0.8 mΩ-cm. This feature was qualitatively and quantitatively similar to that observed for $Bi_4O_4S_3$ in ongoing studies. A broad drop, which was indicative of a superconducting transition, was observed at temperatures below 4 K (see inset), and zero resistivity was confirmed at a temperature of 1.6 K. To confirm the superconducting shielding fraction at the zero-resistivity temperature, the measured pellet was again studied by $\chi$ measurements and the superconducting shielding fraction was confirmed to be smaller than 0.3% even at 1.6 K (Fig. 6). Note that the $\chi$ measurements were conducted on the same pellet used for the $\rho$–$T$ measurements (the pellet was not ground). The extremely small superconducting shielding fraction was thus in sharp contrast to the bulk SC observed for $Bi_4O_4S_3$ in ongoing studies.[1–3] The zero resistivity observed and the trivial superconducting volume fraction (<0.3%) in the $\chi$–$T$ curve provide evidence of the impurity-driven SC of $Bi_4O_4S_3$ synthesized by the HP method at temperatures exceeding 1.6 K. The extremely small-volume SC (<0.3%) showing zero resistivity is likely caused by a trivial superconducting link, which is thinly spread out throughout the grain boundaries and at the surface in a synthesized pellet. Unidentified amorphous thin impurities or altered surfaces of particles of $Bi_4O_4S_3$ are possible causes of the impurity-driven SC observed for the HP-prepared pellet. We investigated the possible impurities.

As the present results appeared to contradict the claim for the discovery of bulk SC in $Bi_4O_4S_3$,[1–3] we attempted to identify any possible contribution by the nonstoichiometry of $Bi_4O_4S_3$ to the superconducting properties. Hence, starting mixtures with S-deficient compositions (10 and 20 at.%) were tested under the optimized synthesis conditions. However, there were no detectable changes in the lattice parameters and magnetic properties (see Fig. 4), indicating that defect structures were not formed in $Bi_4O_4S_3$. Similar results were obtained for O-deficient starting compositions and S-rich starting compositions. Since the sample is chemically stable, as mentioned earlier in the text, decomposition between the time of synthesis and the time of measurement is unlikely to be a cause of the absence of bulk SC. Hence, a possible factor that could account for the impurity-driven SC in the present compound is an undetected difference: for instance, $Bi_4O_4S_3$ prepared at 3 GPa may not be identical to that prepared by the normal process. The local coordination of the structure may be slightly altered in the former case.

To identify the superconducting impurity, polished surfaces of the $Bi_4O_4S_3$ compounds synthesized by the HP method and the normal method were investigated by EPMA. Figures 7(a) and 7(b) show composition images (indicated by CP) and element maps (S, O, and Bi) at the same magnification. The pellet synthesized by the HP method was much denser than the other, reflecting the measured densities of ~6.94 g/cm$^3$ (89% of the calculated density for $Bi_4O_4S_3$) and ~6.11 g/cm$^3$ (78%) for the pellets synthesized by the HP process and normal process, respectively. Moreover, the EPMA results indicated that the bright part is relatively more sulfur deficient and Bi rich than $Bi_4O_4S_3$. Note that the bright part does not correspond to electric charging in the insulating state because it is a composition image. Since the bright parts were confirmed to exist throughout the grain boundaries and at the surface of $Bi_4O_4S_3$ prepared by the normal method, it is reasonable to attribute the superconducting impurity to the bright parts. We thus attempted to identify the parts by XRD analysis and EPMA; however, the studies were unsuccessful probably because the parts were either amorphous or thinly spread out. Perhaps, the extremely low melting points of starting materials (Bi melting point: 272 °C and S melting point: 115.2 °C) may play a key role in forming the bright parts in the heating process; partially melted phases may be included in the chemical reaction paths to form the bright parts.

The present EPMA and XRD data indicated that the various forms of impurities could play a fundamental role in the zero resistivity observed for the HP- and normal-process-prepared samples. Under the synthesis conditions adopted in ongoing studies, the superconducting impurity might form a link unexpectedly in the $Bi_4O_4S_3$ pellet. This link could be responsible for zero resistivity as well as for the full superconducting volume fraction, as was evidenced by the marked decrease in the shielding fraction even upon gentle grinding. The impurity-driven SC in the HP-synthesized $Bi_4O_4S_3$ compound was presumably present because the trivial amount of superconducting impurity (as evidenced by the shielding fraction of the pellet smaller than 0.3%) forms a comparable link. Recent specific heat measurements of a full superconducting compound showed only negligible anomaly at $T_c$, which seemed to support the idea of the impurity-driven SC.[7] Although we were unsuccessful in identifying the superconducting impurity, we could at least conclude that the impurity drove SC with zero resistivity above 1.6 K for the layered oxysulfide $Bi_4O_4S_3$ compound synthesized by the HP method. It is likely that a similar superconducting impurity complicates ongoing studies of the SC of $Bi_4O_4S_3$ prepared by the normal process because its shielding fraction is markedly decreased to less than 1/14 of that of the pellet even upon gentle grinding.

We investigated the superconducting parameters, which were reported for $Bi_4O_4S_3$ in ongoing studies. Magnetic measurements in ref. 3 showed that $H_{c1}$ at 2 K was ~15 Oe and that the lower critical field ($H_{c2}$) at 0 K was 31 kOe. The coherence length ($\xi_{GL}$) at $T = 0$ was then estimated to be ~10 nm, which was nearly comparable to that of $MgB_2$ and much longer than those of Fe-based and cuprate superconductors (typically a few nm).[8] According to the parameters, we could roughly estimate $\lambda_L$ to be ~580 nm using the Bardeen-Cooper-Schrieffer formula $H_{c1}(T) = (\phi_0 / 4\pi\lambda_L^2)\ln(\lambda_L / \xi_{GL})$, where $\phi_0$ is the flux quantum.[9] The $\lambda_L$ was compared with the scale of the pellet, which was used to measure the magnetic properties in this study. The HP-prepared pellet has the minimum scale dimension of ~1 mm, which was much larger than $\lambda_L$; thus, the sample scale unlikely impacted the magnetic measurements. The ground sample for the normally prepared sample has average dimensions of powders of a few 10 μm (confirmed by optical microscopy). Although the average dimension was still much larger than $\lambda_L$, we repeated the same magnetic measurements in a magnetic field of 1 Oe (<< $H_{c1}$) to test a possible impact from the sample scale approaching $\lambda_L$ [Figures 8(a) and 8(b)]. The results, however, showed no notable improvement in the superconducting shielding fraction of the ground sample of the normally prepared sample. Thus, the relative sample scale to $\lambda_L$ hardly accounted for the unusually smaller superconducting shielding fraction of the powder form.

## 4. Conclusions

In summary, we carried out magnetic susceptibility and electrical resistivity measurements as well as electron microprobe analysis on $Bi_4O_4S_3$, which has recently been claimed to be a bulk superconductor ($T_c = 4.5$ K). Quality-improved polycrystalline $Bi_4O_4S_3$ was successfully prepared by an HP method, and the lattice parameters and normal-state $\rho$, as well as the density of states at the Fermi level, were found to be comparable to those determined previously. The most notable discovery in this study was the impurity-driven SC (>1.6 K) in $Bi_4O_4S_3$, which contradicted the results of ongoing studies of this compound. There are two possible reasons for the above-mentioned discrepancy: (i) common technical errors, including those related to the presence of a superconducting impurity, in ongoing studies. The impurity was found to show zero resistivity, even though its superconducting shielding fraction was smaller than 0.3%; (ii) slight local structure and chemical composition differences between the HP phase and the superconducting phase of $Bi_4O_4S_3$, including a small variation of the S1–Bi2 bond in the $BiS_2$ layer, as detected in this study. The concentrations of vacancies of the $SO_4$ group in the layered structure of $Bi_4O_4S_3$ might also vary upon HP heating. If the latter is true, the HP phase can be a host material for the superconducting phase.

Although we cannot discuss the possible structure or composition variations of $Bi_4O_4S_3$ in more detail because neutron diffraction studies are required, we conclude that the SC of the HP-prepared $Bi_4O_4S_3$ is most likely impurity-driven. The relative sample scale to the $\lambda_L$ hardly accounted for the negligible superconducting shielding fraction, as evidenced by the magnetic susceptibility measurements at a magnetic field of 1 Oe as well as of 10 Oe on a mm-size HP sample. Hence, the full shielding fraction reported in ongoing studies[1–3] was possibly complicated by a superconducting link consisting of similar superconducting impurities throughout the grain boundaries and at the surface in a synthesized pellet as evidenced by marked decreases in the shielding fraction at a magnetic field of 1 Oe ($<< H_{c1}$) even upon gentle grinding.

Although the various forms of the superconducting link could play a fundamental role in the observed SC, it was highly challenging to identify the chemical composition and structure of the superconducting impurities because partially melted phases can be essential in the formation of the superconducting link (Bi melting point: 272 °C and S melting point: 115.2 °C). Although the Bi powder treated in the same heating procedure as the reference was not superconducting above 2 K, multiple phases were superconducting at the highest $T_c$ of 8.2 K.[10] Thus, neighbor materials in solid and amorphous forms should be carefully studied for relevance to the observed SC. In addition, the surface of the $BiS_2$ layered phase might be altered under the presence of liquid phases during the heating process, resulting in the appearance of possible surface superconductivity. No matter which, the observed superconductivity did not truly reflect the bulk nature of the $BiS_2$ layered phase, regardless of the manner in which the $Bi_4O_4S_3$ compound is synthesized. Since the difficulty in identifying the superconducting impurities complicates ongoing studies, further experimental efforts toward measuring the degree of impact from impurities and possible altered particle surfaces on the observed SC would be needed to correctly characterize the $BiS_2$ layered phase.

**Acknowledgements**

This study was financially supported by the World Premier International Research Center from MEXT, and by a Grant-in-Aid for Scientific Research (22246083, 25289233) from JSPS, the Funding Program for World-Leading Innovative R&D on Science and Technology (FIRST Program) from JSPS.

**Table I. Structures and isotropic displacement parameters of $Bi_4O_4S_3$ prepared by the HP method.**[a]

| Atom | site | $g$ | $x$ | $y$ | $z$ | $B_{iso}$ (Å$^2$) |
|------|------|-----|-----|-----|-----|-------------------|
| Bi1 | 4$e$ | 1 | 0 | 0 | 0.05785(5) | 0.5 (fixed) |
| Bi2 | 4$e$ | 1 | 0 | 0 | 0.20746(6) | 0.5 (fixed) |
| Bi3 | 4$e$ | 1 | 0 | 0 | 0.38224(5) | 0.5 (fixed) |
| S1 | 4$e$ | 1 | 0 | 0 | 0.1444(3) | 0.5 (fixed) |
| S2 | 4$e$ | 1 | 0 | 0 | 0.2863(3) | 0.5 (fixed) |
| S3 | 2$b$ | 0.5 | 0 | 0 | 0.5 | 0.5 (fixed) |
| O1 | 8$g$ | 1 | 0 | 0.5 | 0.0912(5) | 0.5 (fixed) |
| O2 | 16$n$ | 0.25 | 0 | 0.27101 (fixed) | 0.47051 (fixed) | 0.5 (fixed) |

[a] Note: Space group $I4/mmm$ (No. 139); chemical formula sum: $Bi_6O_6S_{4.5}$ (Z = 2); $a$ = 3.9252(3) Å, $c$ = 41.088(2) Å, $V$ = 633.05(7) Å$^3$; $d_{cal}$ = 7.839 g cm$^{-3}$; $R_{wp}$ = 10.298 %, $R_p$ = 8.002 %. S = 1.967.

**Figure captions**

**Fig. 1.** (Color online) Powder XRD patterns of a series of $Bi_4O_4S_3$ compounds prepared by the HP method. The heating temperatures used for each powder sample are indicated in the figure. The reference pattern for $Bi_4O_4S_3$ sample prepared under AP at 450 ºC is shown at the bottom. Characteristic peaks for the tetragonal $Bi_4O_4S_3$ phase are labeled using Miller indices in the pattern for 700 ºC as a representative. Solid squares indicate the peaks due to impurities. For clarity, commas are used to separate the numbers in some indices.

**Fig. 2.** (Color online) (a) Rietveld analysis of the XRD pattern of the $Bi_4O_4S_3$ compound prepared by the HP method. Markers and solid lines show the observed and calculated profiles, respectively, and the difference is shown at the bottom. The Bragg reflections are denoted by tick marks. (b) Local structure coordination of Bi in the $BiS_2$ layer, drawn on the basis of the present refinement results. The numbers inside are bond distances in Å and angles in degrees. (c) Corresponding local structure view for $Bi_4O_4S_3$ synthesized by a normal method for comparison. The structure data were cited from ref. 1.

**Fig. 3.** (Color online) $T$ dependence of $\chi$ for $Bi_4O_4S_3$ prepared under AP. For comparison, measurements were performed on an as-made pellet and gently ground powder. The largest shielding fraction of approximately 126% was observed for the pellet. The insets show the $T$ dependence of $\rho$ for the as-made pellet of $Bi_4O_4S_3$ (AP).

**Fig. 4.** $T$ dependence of $\chi$ for the compounds prepared under various heating conditions at 3 GPa. The set of curves for each compound includes the ZFC and FC curves measured at a magnetic field of 10 Oe. The data for the Bi reference material (annealed at 700 °C and 3 GPa) are plotted as well. The largest shielding fraction calculated using a material density of 7.84 g/cm$^3$ is 2.2% at most. The inset shows the $T$ dependence of $\chi$ for the single-phase polycrystalline $Bi_4O_4S_3$ under a large magnetic field of 10 kOe.

**Fig. 5.** (Color online) $T$ dependence of $\rho$ for polycrystalline pellet of $Bi_4O_4S_3$ (HP). The inset shows an enlarged view of the low-temperature part (<7 K).

**Fig. 6.** (Color online) $T$ dependence of $\chi$ for gently ground powder of $Bi_4O_4S_3$ prepared under high-pressure conditions of 3 GPa and 700 °C. The ZFC and FC curves were measured under a magnetic field of 10 Oe. Note that the shielding fraction, calculated using a material density of 7.84 g/cm$^3$, is 0.3% at most at the lowest temperature, where zero resistivity was confirmed before grinding. The inset shows an enlarged view of the plots.

**Fig. 7.** (Color online) Comparison of composition image and elemental maps of S, O, and Bi in EMPA for the polished surfaces of $Bi_4O_4S_3$ synthesized under (a) AP and (b) HP conditions.

**Fig. 8.** (Color online) $T$ dependence of $\chi$ for the $Bi_4O_4S_3$ compounds prepared under (a) AP and (b) HP conditions measured at an applied magnetic field of 1 Oe. For comparison, measurements were performed on an as-made pellet and gently ground powder.

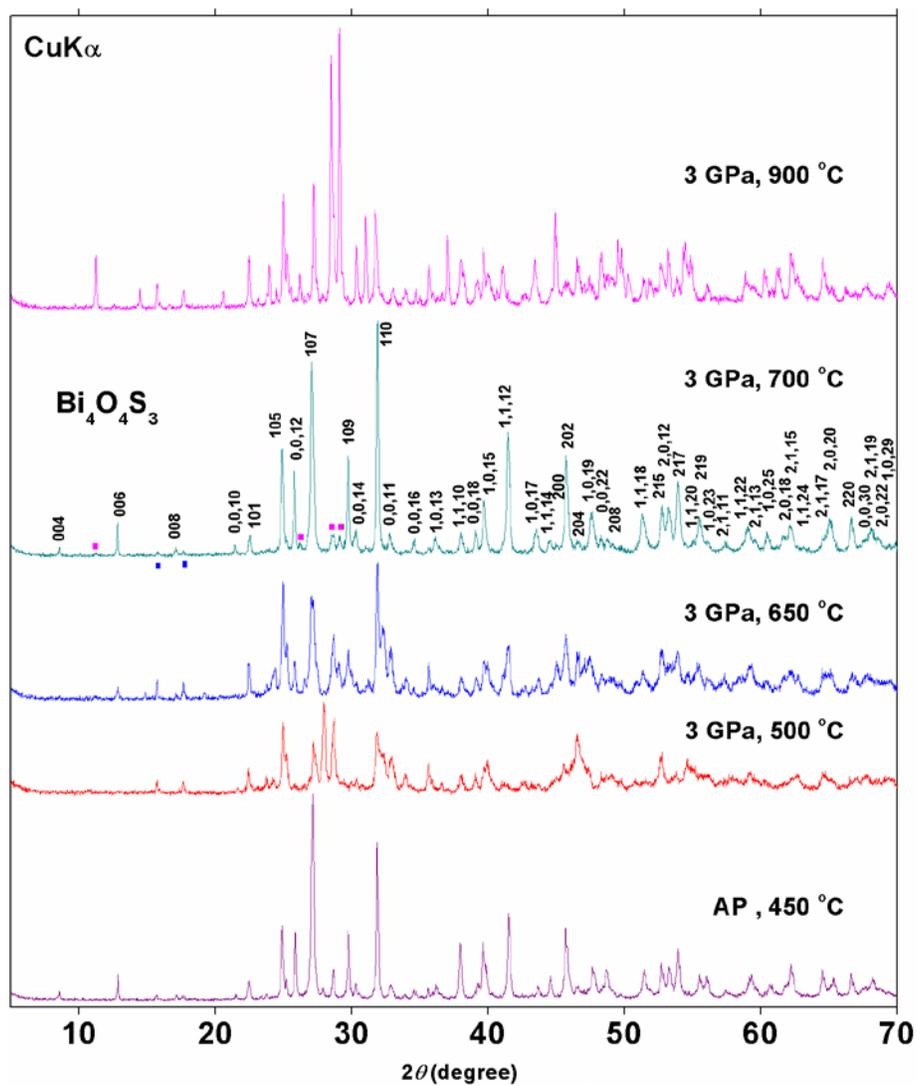

Fig. 1

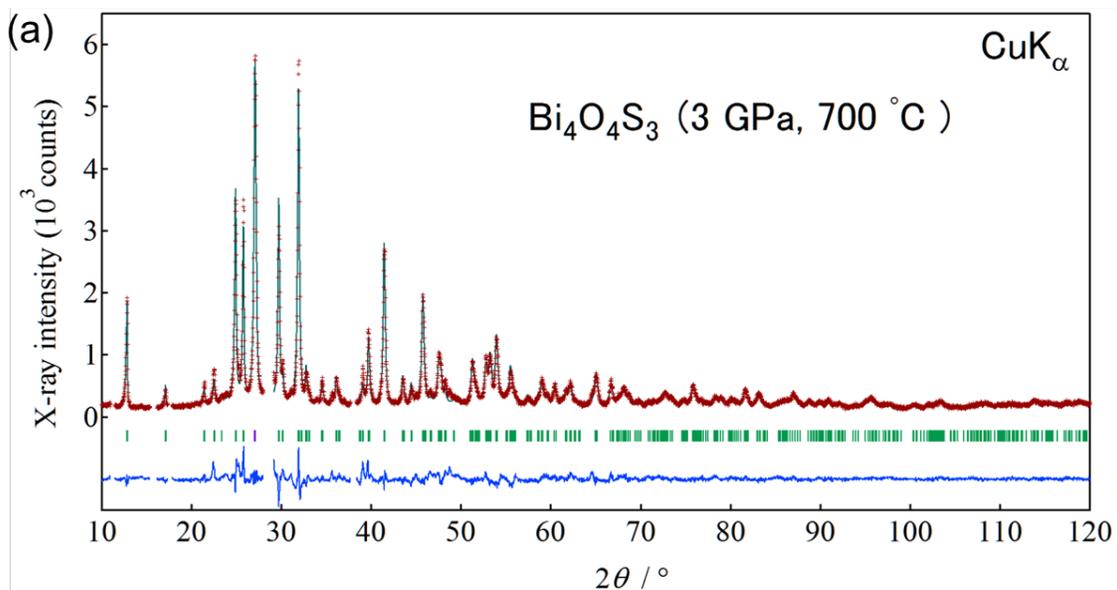
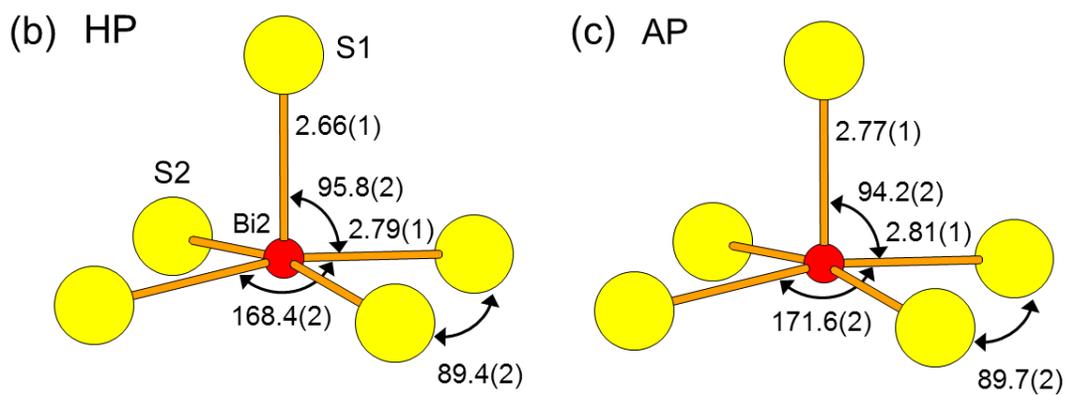

Fig. 2

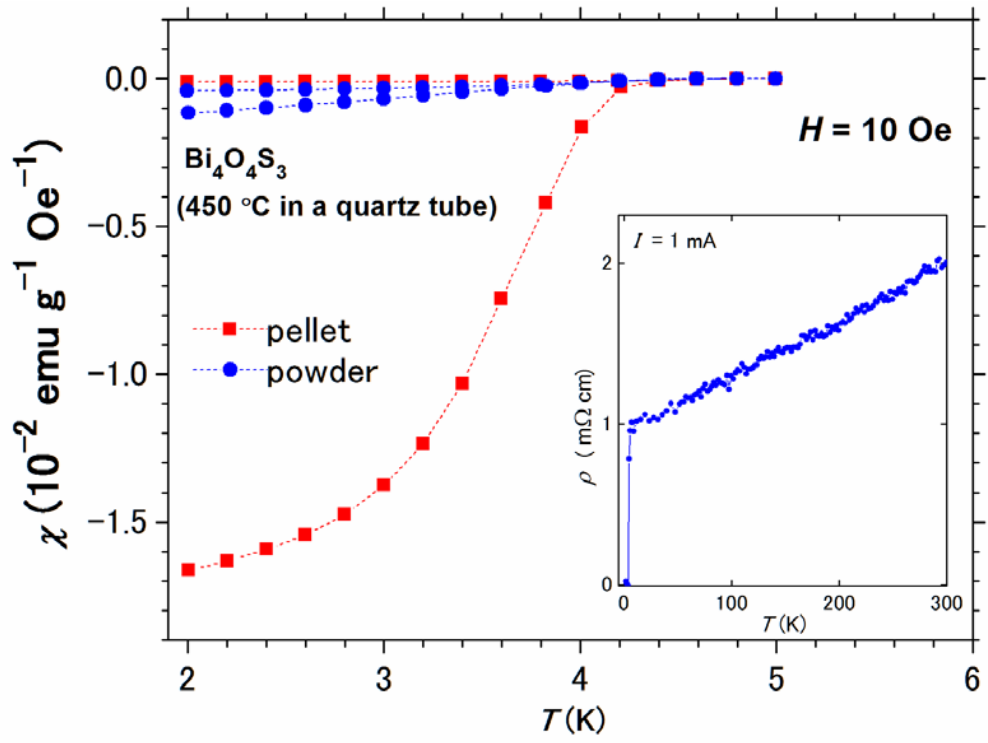

Fig. 3

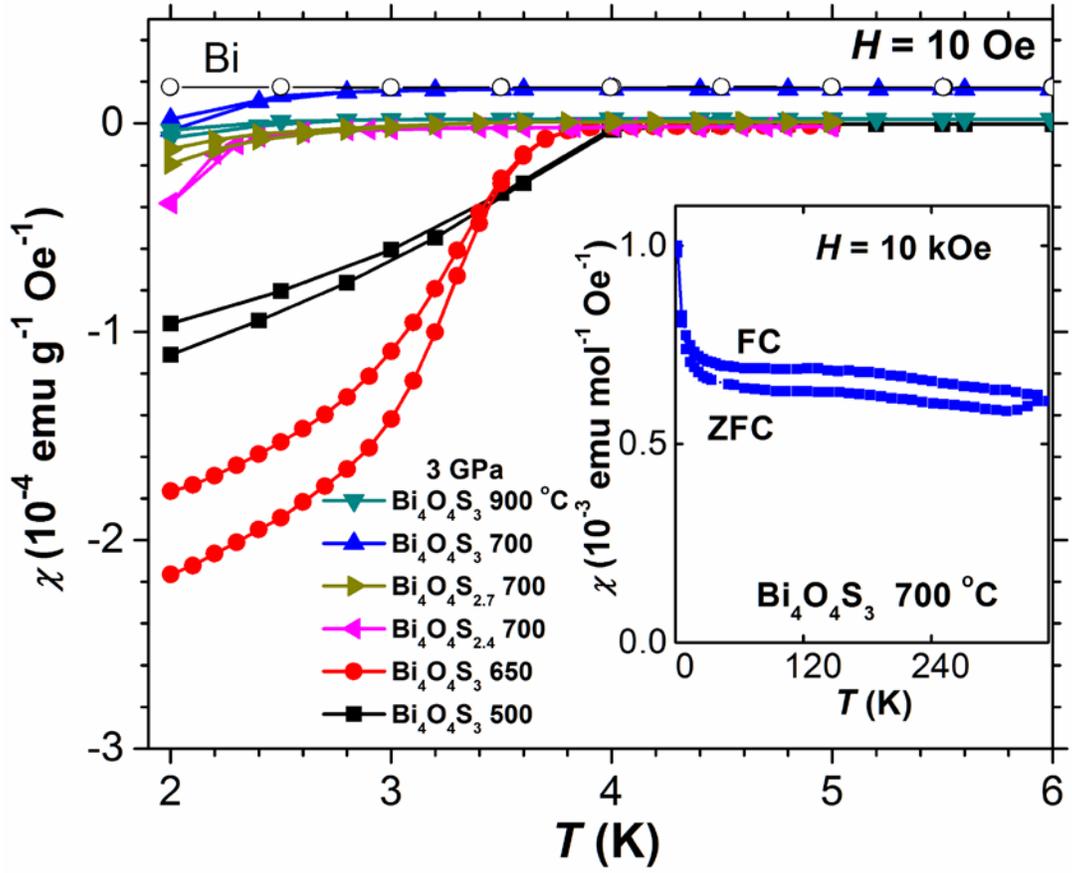

Fig. 4

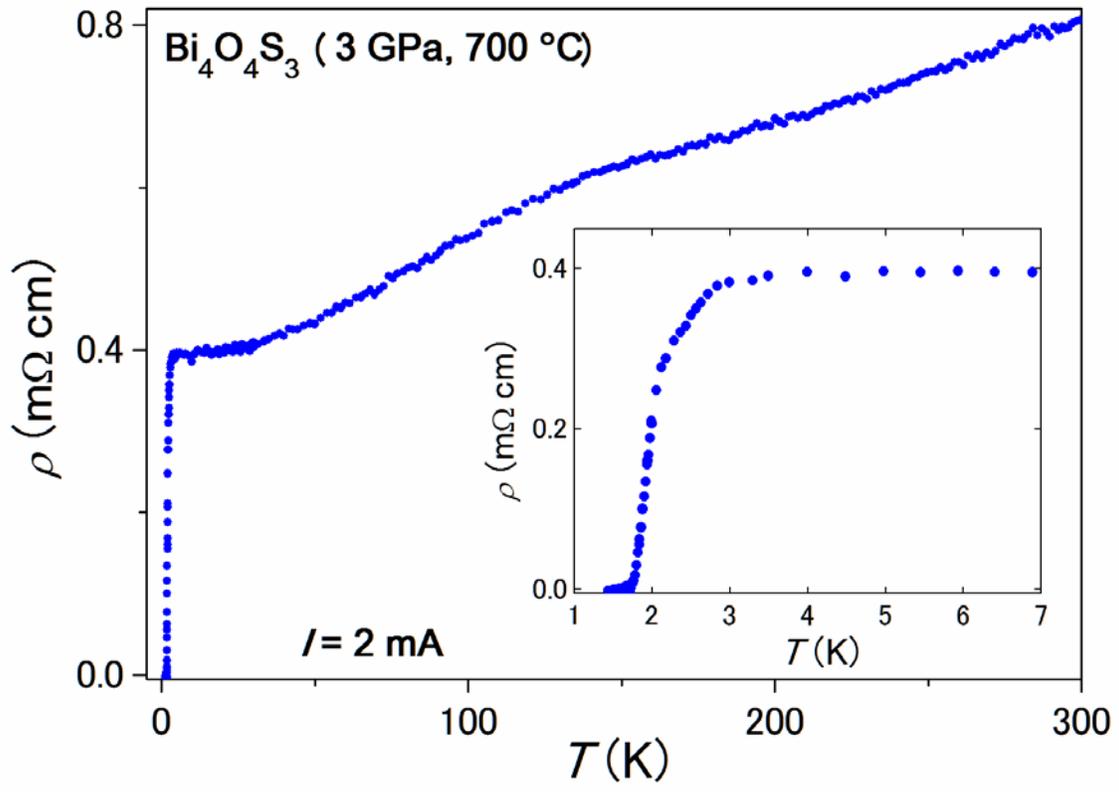

Fig. 5

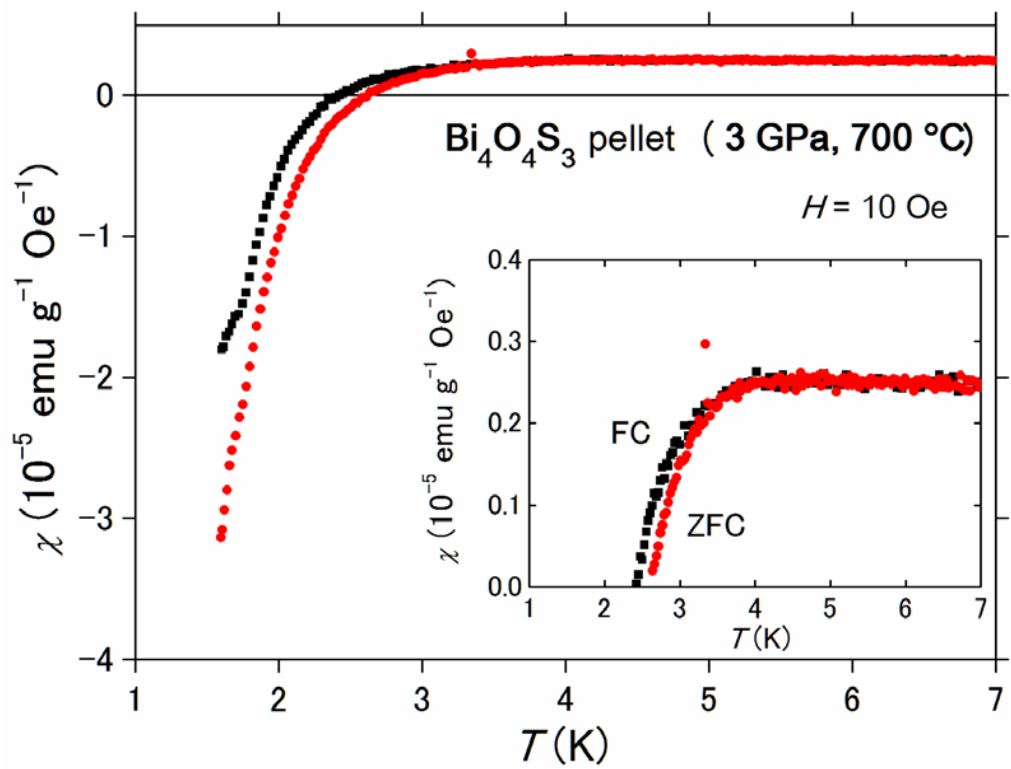

Fig. 6

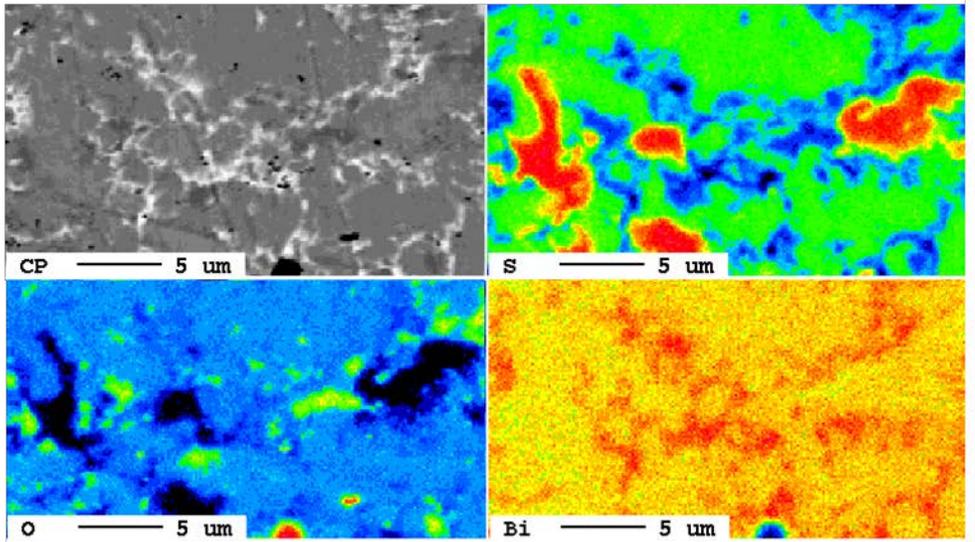

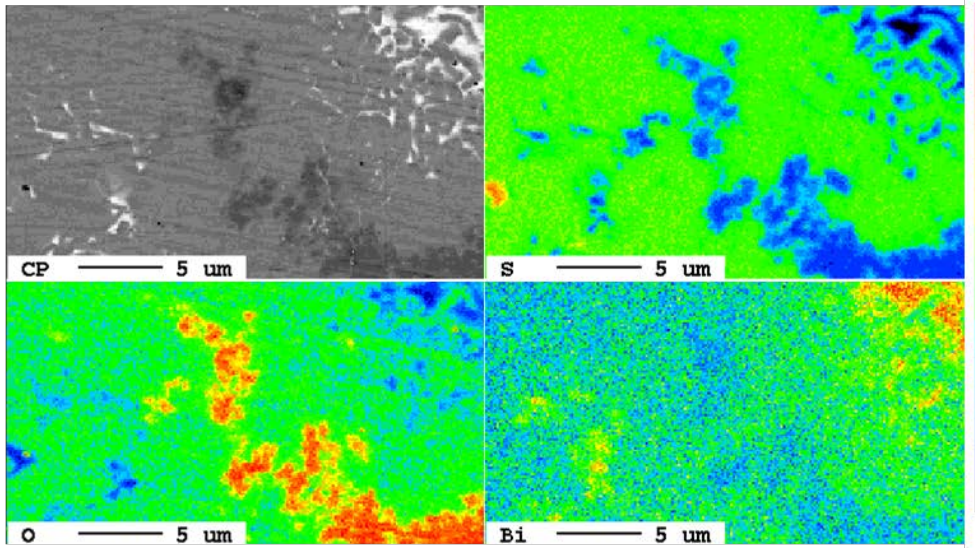

Fig. 7

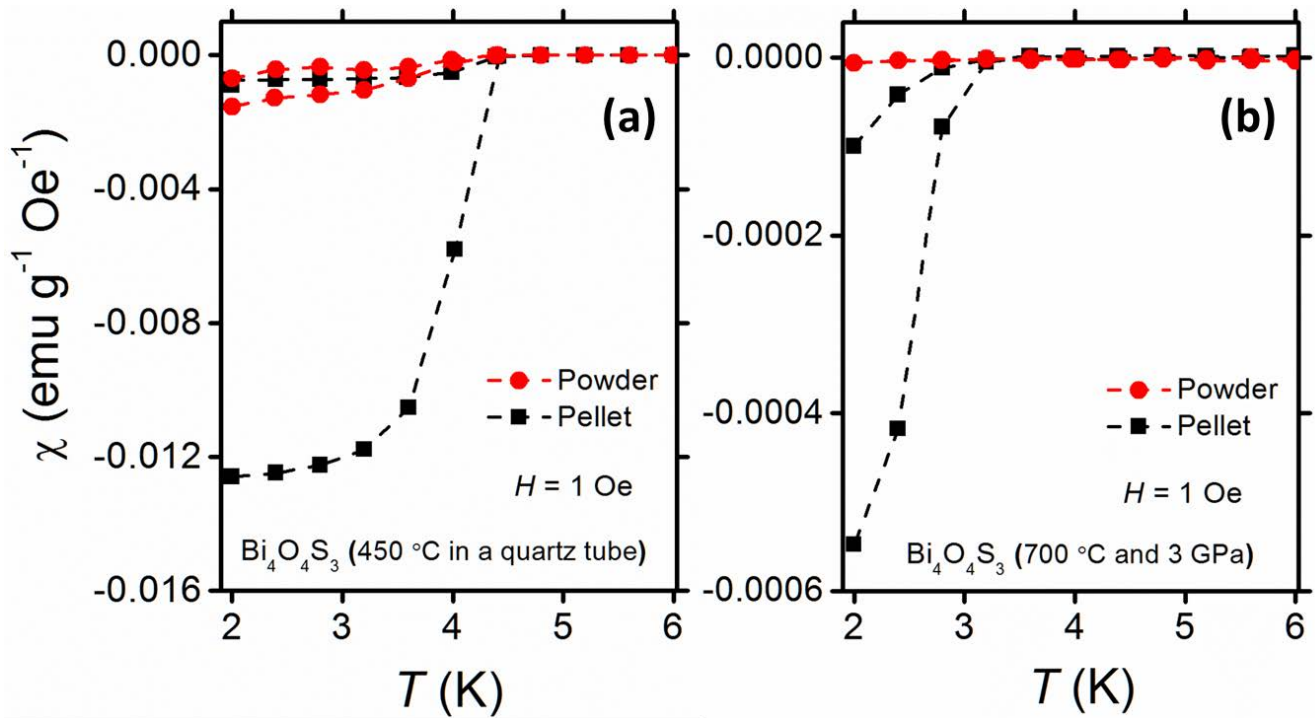

Fig. 8